# SUNSPOT MINIMUM EPOCH BETWEEN SOLAR CYCLES No 23 AND 24. PREDICTION OF SOLAR CYCLE No 24 MAGNITUDE ON THE BASE OF "WALDMEIER 'S RULE"

*Boris Komitov, Peter Duchlev , Konstantin Stoychev, Momchil Dechev and Kostadinka Koleva*

*Abstract* The main purpose of this study is the determination of solar minimum date of the new sunspot cycle No 24. It is provided by using of four types of mean daily data values for the period Jan 01. 2006 – Dec 31. 2009: (1) the solar radioindex *F10.7*; (2) the International sunspot number *Ri*; (3) the total solar irradiance index (*TSI*), and (4) the daily number of X-ray flares of classes from "B" to "X" from the soft X-ray GOES satellite channel (0.1 – 0.8 nm). It is found that the mean starting moment of the upward solar activity tendency (the mean solar minimum) is Nov. 06$^{th}$, 2008. So, the solar cycle No 23 length is estimated to ~12.6 years. A conclusion for a relatively weak general magnitude of solar cycle No 24 is made. By using of relationship based on the "Waldmeier's rule" a near maximal mean yearly sunspot number value of 72 ± 27 has been determined.

*1. Introduction*

There are clear evidences that the last solar minimum (AD 2006-2009) is the deepest one since AD 1913, when the minimum between the Schwabe-Wolf's sunspot Zurich numbered cycles 14 and 15 occurred. As a confirmation one could point out that:

   1. There had been about 700 spotless days since Jan 26$^{th}$ , 2004 up to Dec 31th, 2009;

   2. In the middle of 2008 the *F10.7* radio-index reached extreme low levels for the whole period of observations since AD 1947 (the mean monthly value for July 2008 was 65.7).

   3. The lowest solar wind parameters for the whole period of observations (about 50 years);

   4. The smallest number of radio bursts in 2008 and 2009 in the MHz–range for the whole period of regular observations (since AD 1960).

   5. The almost total absence of geomagnetic activity during 2008 and 2009.

The first magnetic active region, related to the new Zurich cycle No 24 was observed in Dec 13$^{th}$, 2007, while the first official sunspot group (NOAA 11980) in Jan 4$^{th}$, 2009. However, these events were followed by a very long period of almost 20 months until Sept 2009, with almost no activity manifestations. A stable slow increase of the new solar cycle began in Sept 2009. This is probably a sign that the minimum between the both cycles is already passed.

The problem for the official starting point of cycle No 24 and the length of the old 23$^{rd}$ is still the topic of the day. This paper presents new results for the onset of cycle No 24.

*2. Data and methods*

The mean diurnal values of four solar activity indexes for the period Jan 1$^{st}$, 2006 – Feb 28$^{th}$, 2010 or total 1521 days are used in this study. These data contains the final downward phase of cycle No 23 up to Apr 2008, the deepest "quiet" phase (Apr 2008- Sept 2009), and the initial significant solar activity increasing after Sept. 2009. We use selected data for:

A. The international daily sunspot number;
B. The radioindex *F10.7*;
C. The daily numbers of X-ray flares (total for classes B, C, M, and X from the soft X-ray GOES satellite 0.1 – 0.8 nm channel);
D. The total solar irradiance (*TSI*-index);
E. The mean monthly sunspot numbers *Ri* for solar cycles No 7 –23 during the first 18 months after the corresponding minima for prediction of the amplitude of solar cycle No 24.

Thus, our study is based on data, related to solar activity phenomena in different regions of solar atmosphere. The data for series A, B, C, and E are taken from the National Geophysical Data Center (*ftp://ftp.ngdc.noaa.gov/STP*), while those for *TSI* – from the SORCE satellite web-site (http://lasp.colorado.edu/sorce/).

The conventional method for detection of the minimal sunspot activity amplitude is based on 13-month smoothed sunspot numbers. The fact that the increasing phase (only 3-4 months up to Feb 2010) is very short suggests that this method could be not sensitive enough on this stage.

A least square procedure for obtaining parabolic full quadratic and qubic polynomial minimized functions of the type $\psi = at^2+bt+c$ or $\psi = at^3+bt^2+ct+d$ are used over each one of the studied data series. There $t$ is the number of days since Jan $1^{st}$, 2006. The both types of trend polynomials for each series has been compared and the better approximation has been chosen by Snedekor-Fisher's *F*-parameter. Thus, the general trends are expressed by simple non-linear functions, in which only one extreme (minimum) in the studying interval exists. There could be obtained the starting moments of cycle No 24 for every one of these data series searching for the minima of the corresponding mean least square minimized polynomials.

*3. The solar cycle No 24 minimum*

The calculated minima of the best trend functions are shown in Table 1 where the minima are given in calendar dates. The best-expressed trends are for the sunspot *Ri* series, as well as for *F10.7* where the corresponding coefficient of correlations *R* are 0.53 and 0.66. Both they belong to cubic type. The trend function for *TSI* correlates to the original data satisfactorily well (*R*=0.49), while this R value very close both to the quadratic and cubic polynomial trend type approximations. For the X-flare events it is *R*=0.38 at cubic approximation. It is important to note that because of the large number of data (1521) all these values of *R* are statistically significant over 95%. The better cubic trend approximation as the quadratic one for Ri and F10.7 data is caused by asymmetry effect of the faster increasing of these indices during the initial upward phase of solar cycle 24 as the decreasing during the downward phase of the solar cycle No 23 before the minima. This asymmetry is no observable in the *TSI* dynamics during the same time.

Table 1
The calculated Zurich cycle No 24 minimum calendar dates

| Solar index | Calendar date | Half interval error [days] |
|---|---|---|
| *Ri* | Nov 14, 2008 | 54 |
| *F10.7* | Oct 28, 2008 | 31 |
| *Soft X-ray flares* | Nov 06, 2008 | 196 |
| *TSI* | July 04, 2008 | 77 |

As it is afore shown the most reliable solar cycle minima estimations are on the base of *F107* and *Ri* data series, where the half interval error is near or less than two months (54 and 31 days, respectively). The uncertainty of the solar cycle minimum on the base of estimations over X-flares and *TSI* series is much higher (±3-6 months with respect to the corresponding calendar dates). The earliest is the TSI–minimum - in the middle of 2008. The slight increase of TSI in the second half of 2008 could be related to the corresponding increase of the total area of the bright regions (the faculae regions and the so-called "ephemeral regions" (ER)) at this time. Such overtaking of the bright ER minima to the main sunspot one for the last Schwabe-Wolf's cycles is discussed by Krivova et al. (2007).

The mean minimum moment for the solar cycle No 24, based on the results in Table 1 (*Ri* and *F10.7*) is Nov 06$^{th}$, 2008. It coincided with the X-ray flares minimum. On the other hand, it is close to the results of Didkovsky et al. (2009). Thus, for the length of the Zurich cycle No 23 started in May 1996 we could obtain ~12.6 years.

It is interesting also to compare the mean activity level of cycle No 24 during the first year after its preceding minimum to the corresponding ones of the previous few cycles. The mean monthly numbers of *Ri* and *F10.7* during their first year after minima for Zurich cycles 19-24 are shown in Table 2.

If the mean monthly values of these indices during the first year after minimum is taken as a criteria for their starting increase, the cycle No 24 should be the weakest for all last six ones both in *Ri* and *F10.7* series. These criteria could be taken as a pre-indicator for a very low near-maximal magnitude of this cycle, which is in good accordance with some other predictions. Such assumption is in the direction of many other forecasts for low amplitude of cycle No 24, as well as for starting of new Dalton-type supercenturial solar minimum (Badalyan, Obridko, and Sykora, 2000; Komitov and Bonev, 2001; Komitov and Kaftan, 2003; Schatten and Tobiska, 2003; Ogurtsov, 2005; Clilverd et al., 2006; Komitov, 2007). In our opinion, the presented results in Table 2 should not be considered as strong evidence, but only as a possible scenario. This is related to the fact that the trends in Table 2 are obviously not very strong. The strongest cycle No 19 (maximal annual *Ri*=190 in 1957) has a very moderate behavior during its first year. On the other hand it is interesting to note that the "starting year" of cycle No 22 (with a maximum in 1989-1990) was characterized by the highest *Ri* and *F10.7* in comparison with all others. This could be a possible indication that solar cycle No 22 was "critical" for the long-term solar activity dynamics and it mark the "smoothing" peak of the unusually high solar activity epoch from 1940 to 1996 or 2000, which is called at last time as "Modern supercenturial solar maximum".

Table 2
The averaged monthly *Ri* and *F10.7* values during the starting
first years of solar cycles No 19-24

| Cycle No | *Ri* (first year) | *F10.7* (first year) |
|---|---|---|
| 19 | 7.5 | 72.7 |
| 20 | 14.2 | 70.0 |
| 21 | 13.2 | 73.5 |
| 22 | 23.4 | 80.3 |
| 23 | 8.8 | 73.2 |
| 24 | 2.2 | 69.9 |

**4.** *Prediction of the solar cycle No 24* a*mplitude*

The next task of our study is the estimation of the near maximum value of the annual sunspot number ($W_{max}$) for solar cycle No 24, based on so-called "Waldmeier's rule" (Waldemeier, 1935). That is expressed by the tendency of the increasing sunspot number rate during the upward phase of the Schwabe-Wolf's cycles, which is higher for the more powerful cycles than for the weaker ones.

Up to this moment (June 2010) there are only less than 2 years after the start of cycle No 24. By this reason we use information for the sunspot activity rising during the initial 18 months of all solar cycles after AD 1818, i.e. No 7-23. Using the corresponding mean monthly sunspot numbers $Ri$, we calculate the mean monthly increasing $\alpha = dRi/dt$ for each solar cycle. The linear correlation relationship between $\alpha$ and $W_{max}$ has been found (fig.1). The coefficient of correlation R is +0.69 and the relationship is:

$$W_{max} = 2.44\alpha + 69.7 \qquad (1)$$

We calculate the parameter $\alpha$ for the first 18 months after the minimum preceding solar cycle No 24. The result is $\alpha = 0.91$. Thus, on the base of equation (1) it is found that the magnitude of this cycle should be $W_{max} = 72 \pm 27$. The obtained value is much less than the magnitude of previous cycle No 23 (~121) and it is also an other indication about a oncoming Dalton-type long term solar minimum. It is interesting to point out that according a new prediction of Kane (2010), based on Ohl's method, $W_{max}$ should be even less (57 ± 25).

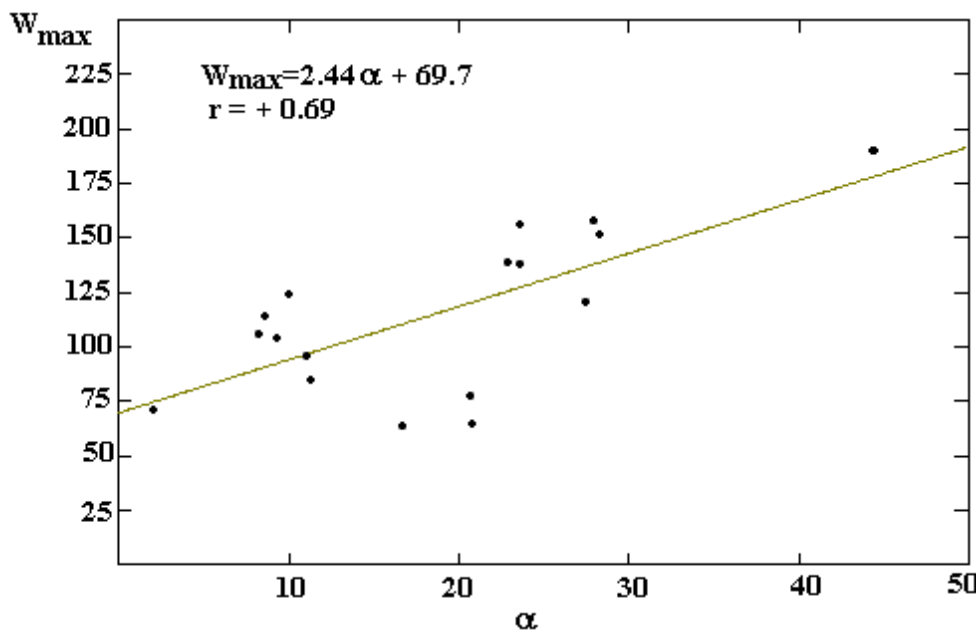

*Fig.1 The relationship between $\alpha$ and Wmax for solar cycles No 7-23*

*5. Conclusions*

**On the base of the analysis presented in this study, one could conclude that:**

1. **The local smoothed minimum between the solar Schwabe-Wolf's cycles No 23 and 24 occurred at the end of AD 2008. The averaged moment of this event, calculated on the base of four solar indexes ($Ri$, $F10.7$, X-ray flares and $TSI$) is Nov 06, 2008.**

2. **The length of solar cycle No 23 is estimated of ~12.6 years.**

3. The comparison between the mean "first year" values of *Ri* and *F10.7* for the Zurich cycles No 19 - 24 shows that the start of cycle No 24 is the weakest of the last six cycles. It could be also an indicator for its weak general magnitude.

4. By using of relationship based on the "Waldmeier's rule" a near maximal value of mean yearly sunspot number of 72 ± 27 has been obtained for the sunspot cycle No 24.

*Acknowledgements.* The authors are thankful to the National Geophysical Data Center, Boulder, Colorado, U.S.A. for providing data for the International sunspot number, radioindex *F10.7*, daily numbers of X-ray flares, and mean monthly sunspot numbers via *ftp://ftp.ngdc.noaa.gov/STP*, and the SORCE satellite team for providing TSI data via http://lasp.colorado.edu/sorce/.
.

Institute of Astronomy
Bulgarian Academy of Sciences
72, Tsarigradsko chausée
1784 Sofia, Bulgaria
e-mail: b_komitov@sz.inetg.bg;